\documentclass[cits]{PoS}

\title{
\rule{0cm}{2.5cm}\vspace{-3.5cm}\\{ 
\it \normalsize IPPP/10/50 \qquad DCPT/10/100}
\vspace{2cm}\\
Automation of Multi-leg One-loop virtual Amplitudes}

\ShortTitle{Automation of 1-Loop Amplitudes}

\author{\speaker{Daniel Ma\^{\i}tre}\\
        Institute for Particle Physics Phenomenology\\
        University of Durham\\
		Science Laboratories\\
		South Rd\\
		DURHAM DH1 3LE, UK
        E-mail: \email{daniel.maitre@durham.ac.uk}}

\abstract{In the last few years, much 
progress has been made in the computation of one-loop virtual matrix elements for processes involving many external particles. In this contribution the methods that have enabled  this recent progress are briefly reviewed with a focus on their computing and automation aspects.}

\FullConference{13th International Workshop on Advanced Computing and Analysis Techniques in Physics Research, ACAT2010\\
		February 22-27, 2010\\
		Jaipur, India}

\def\Zgamjjj{$Z,\gamma^*\,\!+\,3$}

\begin{document}

\section{Introduction}
Theory predictions play a very important role in the particle physics experiments at current hadron colliders. They are not only needed for the signal, that one aims at measuring, but also are required for the modelling of the many backgrounds that make most experimental measurements very challenging. In many cases the signal-to-background ratio is close to or below unity, so that a valuable measurement is only possible with a reliable understanding of the background processes.
\section{NLO corrections}
The first step for a theoretical prediction is a tree-level calculation. This type of computation is highly automated and many programs exist \cite{Alpgen,Amegic,Sherpa,compHEP,Helac,Herwig,Madgraph,Pythia,Whizard}. Unfortunately these predictions suffer from a large dependence on the unphysical factorization and renormalisation scales. This dependence is due to the truncation of the perturbation series. At tree level this dependence is monotonic and therefore tree-level predictions cannot provide a quantitative prediction for the absolute normalization of cross sections. The problem gets worse with each power of the coupling constant added as one increases the number of jets in the process. This fact is illustrated in Table \ref{XSecTable}.
\begin{table}[th]
\begin{tabular}{|c|c|c|c|c|}
\hline \# of jets & LO & \% scale dep. & NLO & \% scale dep. \\ 
\hline 1 & $4635^{+928}_{-715}$ & 20\% & $6080^{+354}_{-402}$ & 6\%\\ 
\hline 2 & $429.8^{+171.7}_{-111.4}$ & 40\% & $564^{+59}_{-70}$  & 12\% \\ 
\hline 3 & $24.6^{+14.5}_{-8.2}$ & 59\% & $35.9^{+7.8}_{-7.2}$ & 21\% \\ 
\hline 
\end{tabular}  
\caption{Cross sections with scale dependence for \Zgamjjj jets. The numbers are taken from Table II of Ref. \cite{Z3jTev} for the cross section of a Z boson or off-shell photon accompanied by up to three jets. The scale dependence is evaluated by multiplying the scale by factors of two and one half compared to the central scale, which is chosen on an event-by-event basis to be the global observable $H_T=\sum_p E^p_T+E_T^{e^+}+E_T^{e^-}$, see \cite{Z3jTev} for the exact definition. The jet algorithm used is SISCone \cite{SISCone}. A complete description of the cuts used can be found in Ref. \cite{Z3jTev}. The integration errors are not displayed as they are much smaller than the scale dependence.
}\label{XSecTable}
\end{table}

Next-to-leading order (NLO) is the first order at which the scale dependence of the coupling constant is partly counterbalanced by the scale dependence of the matrix elements. Therefore NLO is the first order at which a quantitatively reliable prediction can be provided. NLO corrections are typically large and can, in addition, affect shapes of distributions significantly. 

To compute NLO corrections, one has to compute two different pieces in addition to the tree-level calculation. The two parts, called the real and the virtual parts can be computed independently but they are both separately divergent. When combined the divergences cancel, yielding a physically meaningful result. 
\subsection{Real part}
The real part of the NLO corrections accounts for the emission of an additional parton into the final state. The real part is infrared divergent when the emitted particle is either collinear to another final state parton or is soft. To allow for the numerical evaluation of the phase-space integration of the real part, its divergences have to be regulated. The most common way to do so is the subtraction method. In this approach, an approximation to the real matrix elements is constructed in such a way that the approximation reproduces the divergent behaviour of the real matrix element in all singular limits. In addition the approximation should be easy enough to integrate analytically over the divergent phase-space, thereby exposing the divergences explicitly which allows the direct cancellation with the explicit divergences of the virtual part. Different methods have been developed to construct these approximations \cite{FKS,CS,Antenna} and several automated programs can provide this approximated matrix elements \cite{AutomatedSubtraction}.    
\subsection{Virtual part}
The virtual part is the current bottleneck towards a completely automated program for NLO cross sections. The standard method to compute the virtual corrections involves computing all Feynman diagrams associated with the process and applying the Feynman rules. This results, for complicated processes, in a very large number of tensor integrals. Each of these tensor integrals can be written in terms of scalar integrals using Passarino-Veltman reduction. This step usually involves solving very large systems of equations and results in very large and potentially numerically unstable expressions for the coefficients of the scalar integrals. The final answer for the one-loop amplitudes takes the form 
\begin{equation}\label{OneLoopDecomposition}
{\cal A} = R+\sum d_{i} I^4_i+\sum c_{i} I^3_i+\sum b_{i} I^2_i,
\end{equation}
where $I^n$ is an $n$-point scalar integral and the sums run over all possible configurations. The scalar integrals are well known and public libraries for their evaluation are available \cite{IntegralsExplicit}. $R$ is the so-called rational part that does not contain any logarithm or polylogarithms. For simplicity we will assume that the internal propagators are massless and therefore no tadpole integrals are present. 
 
Despite these difficulties the standard Feynman diagram approach has recently been applied to compute challenging state-of-the-art processes, such as the production of a top-quark pair and a bottom-quark pair \cite{BDDP} or the production of two bottom-quark pairs \cite{bbbb}. 

Recently, a new set of approaches have emerged that take advantage of the knowledge that the final answer can be written in terms of a basis of scalar integrals and aim at computing their coefficient directly avoiding any computationally intensive integral reduction. 
\subsection{The Ossola-Papadopoulos-Pittau method}
Reduction at the integrand level was first proposed in Refs. \cite{delAguila,OPP}. Here we only attempt to sketch the method, for a complete description, the reader is referred to Ref. \cite{OPP}. The idea is to identify at the integrand level the structures that will generate the scalar integrals and their coefficients and those that will vanish upon integration of the loop momentum. 

The starting point is an ansatz for the one-loop amplitude
\begin{equation}\label{OPPAnsatz}
{\cal A}=\int dl \frac{{\cal N}(l)}{\prod_j D_j}\qquad ,\qquad {\cal N}(l)=\sum c_i T_i(l)\,,
\end{equation}
where the $D_j$'s are the propagators that can occur and the structures $T_i(l)$, when integrated over the loop momentum, are either scalar integrals or zero.
\begin{equation}
\int dl \frac{T_i(l)}{\prod_j D_j}=\left\lbrace\begin{array}{c} 0\\ S_j\end{array} \right.\,,
\end{equation}
where $S_j$ are scalar one-loop integrals. The key for the OPP method is to identify all possible structures in the numerator, not only those that generate scalar integrals so that the full numerator can be accounted for. The strategy is now to find the coefficients $c_i$ in Eq.~(\ref{OPPAnsatz}) by evaluating the numerator ${\cal N}(l)$ for several different values of the loop momentum $l$. It would be impractical to attempt to solve for all $c_i$'s at the same time, as the system of equation would be much too large for an efficient numerical technique. Instead, the OPP method proposes to use selected momenta for which most of the structures $T_i$ vanish. These correspond to momenta for which a given set of inverse propagators vanish, which is a connection between the OPP method and the generalized unitarity. This is one of the  In this way the size of the system of equations to solve stays manageable and all coefficients $c_i$ can be numerically computed. The full one-loop amplitude is recovered by multiplying the coefficients with their corresponding scalar amplitudes.

The OPP method has been successfully applied to a large number of processes, see Ref. \cite{BOPP,HPP,Actis:2009uq,ttbbCzakon,ttjj}. A public program called {\sc CutTools} that computes the coefficients $c_i$ in Eq.~(\ref{OPPAnsatz}) given the numerator function is publicly available \cite{CutTools}. 

The OPP method and the unitarity-based methods (to be described below) are closely related and can be combined \cite{EGK}. Very recently a new public program called {\sc SAMURAI} that extends the OPP method to compute the rational part directly in the OPP framework using D-dimensional unitarity \cite{GKM} has been made available \cite{Samurai}.
\subsection{Unitarity-based methods}
Another set of methods exploiting the fact that a one-loop amplitude can be written as a sum of coefficients multiplying scalar integrals are the generalized unitarity methods \cite{UnitarityMethod}. These methods use the general factorization properties of the amplitude as a tool to compute coefficients of the scalar amplitudes in Eq. (\ref{OneLoopDecomposition}). 

This section sketches the unitarity-based methods, for a complete review, see Refs.~\cite{OnShellReviews}. The unitarity-based methods use so-called generalized unitarity cuts which act to replace the loop integral of multiple propagators with delta-functions
\begin{equation}
\frac{1}{P^2}\rightarrow 2\pi i \delta(P^2)\,.
\end{equation}
Unitarity cuts can be seen as projectors that project onto everything that contains all the propagators of the cut. 

The unitarity method works by applying a set of unitarity cuts to both sides of Eq.~(\ref{OneLoopDecomposition}). Quadruple cuts were first applied in Ref. \cite{BCFUnitarity}. If we apply a quadruple cut on the right-hand side of Eq.~(\ref{OneLoopDecomposition}), the cut operation will single out the coefficient of the single box scalar integral that has the four propagators of the cut. What is the effect of the quadruple cut on the left-hand side of Eq.~(\ref{OneLoopDecomposition})? The effect of a quadruple cut in four dimensions is to freeze the loop integral, since four denominators are transformed into delta-functions, all integration variables in the integrand are fixed by the four cut conditions. In addition, the factorization properties of the amplitude ensure that the integrand factorises into a product of the four tree-level amplitudes that are singled out at the four corners of the quadruple cut, with their external legs along the cut propagators evaluated at a value of the loop momentum that satisfies the four cut conditions:     
\begin{equation}
\parbox{0.1\textwidth}{\includegraphics[scale=0.3]{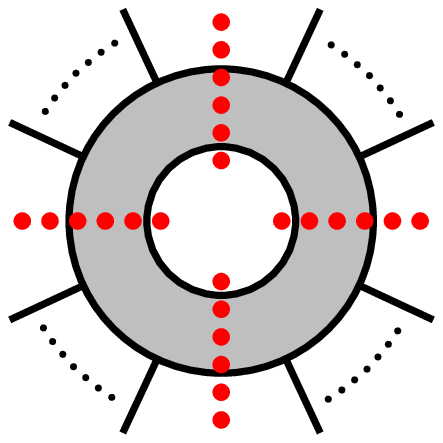}}=\sum\limits_{l}\, A_1^i(l)A_2^i(l)A_3^i(l)A_4^i(l)
\end{equation}
where the sum is over all solutions of the loop momenta that satisfy the four cut conditions. The advantage of expressions derived in this way compared to those obtained by the traditional reduction methods, beyond the simplicity of their calculation, is that they are much more compact. This is partly owed to the fact that all ingredients are on-shell tree amplitudes and therefore don't carry any gauge information which usually clutters the coefficients obtained by the standard method while cancelling in the final answer. Such compact expression are numerically very stable.

In a triple cut one chooses three propagators to promote to delta-functions. Since the loop integration is four-dimensional and we have three conditions imposed by the cut, we will be left with a one-dimensional integral. The left-hand side of Eq.~(\ref{OneLoopDecomposition}) will have the schematic form
\begin{equation}
\parbox{0.1\textwidth}{\includegraphics[scale=0.3]{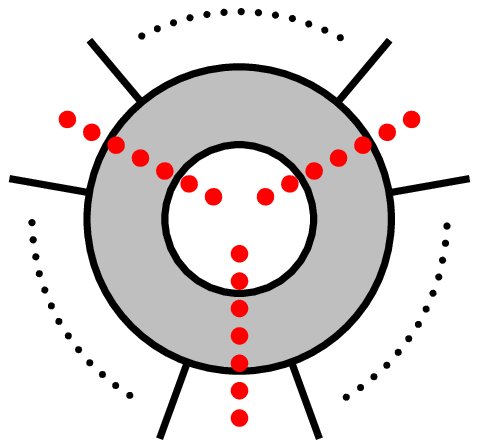}}=\int dt J(t)A_1(t)A_2(t)A_3(t)
\end{equation}
after a triple cut has been performed. Again the integrand splits up into a product of tree amplitudes due to the factorization properties of the one-loop amplitude. The right-hand side of Eq.~(\ref{OneLoopDecomposition}) will get two types of contributions when a triple cut is applied. There will be a contribution containing the coefficient of the triangle scalar integral that has exactly the three propagators chosen for the cut and also contributions from box scalar integrals that have the three propagators chosen for the triple cut, along with an additional uncut one. This additional propagator contains a pole in the remaining free loop momentum parameter. So with a proper choice of parametrization for the loop momentum and the knowledge of the parametric form of the integrand in this parametrization it is possible to disentangle the triangle coefficient and the box contributions in the triple cut \cite{Forde}. In a numerical approach one can take advantage of the analytic form of the integrand on the complex plane to subtract the box poles, as these can be computed as outlined above, from the triple cut and compute the triangle coefficient~\cite{BlackHatI}.

A similar strategy can be used to compute the bubble coefficients. We refer to \cite{Forde,BlackHatI} for more details. 

The only part of the one-loop decomposition Eq.~(\ref{OneLoopDecomposition}) that cannot be computed using four-dimensional unitarity is the rational part $R$. Several methods exist to compute this last piece. 
\begin{description}
\item[{\it  Special Feynman rules}] One can use specially crafted Feynman rules to compute only the rational part of the one-loop amplitude \cite{XYZ} or the part of the rational term that cannot be inferred from the cut containing part \cite{OPPR2}. 
\item[{\it Recurrence relations}] One can also use a generalization of the tree-level BCFW recursion \cite{BCFW} for the case of the rational part of the one-loop amplitude \cite{Bootstrap}. This strategy has also been implemented numerically in the {\sc BlackHat} program \cite{BlackHatI}.
\item[{\it D-dimensional Unitarity}] Alternatively, one can perform the cuts in higher dimensions and use the additional information to compute the rational part. Several strategies can be employed. First one can compute the cuts in two different integer dimensions and combine the information to get the rational part \cite{GKM}. This strategy has been employed in the {\sc ROCKET} program \cite{EMZ,EMZ2,ggttg}. One can also relate the additional $D-4$ components of the loop momentum to massive terms in four dimensions \cite{DdimUnitarity,Badger}. A numerical version of this approach has been used in the {\sc BlackHat} program \cite{W3j,W4j}.    
\end{description}

The unitarity-based methods have recently lead to state-of-the-art phenomenology results for important processes for the LHC, such as W+3 jets \cite{W3j,EMZ,EMZ2,MZW3jets}, Z+3 jets \cite{Z3jTev} and progress towards W+4 jets \cite{W4j}. Alongside these numerical programs, compact analytic expressions have been found using unitarity-based techniques for the virtual matrix elements for the hadroproduction of a Higgs boson accompanied by two jets \cite{Hjj}.

\section{Automation}
The computation of the real part of NLO corrections has already been achieved, with several programs \cite{AutomatedSubtraction} providing the necessary ingredients. The current bottleneck for the automation is the automatic computation of virtual one-loop amplitudes. Most ingredients for the NLO corrections are "tree-like" and are implemented in a larger software framework that can deal easily with tree amplitudes such as, for example, Helac, Herwig, Madgraph {\textit MadEvent?} or Sherpa. Typically the virtual amplitudes needed to complete the NLO computation are provided by a seperate program. In order to simplify the interface between the programs computing virtual one-loop amplitudes and "client" frameworks, a standard interface has been drafted during the Les Houches 2009 workshop. 

\subsection{Binoth Les Houches Accord}

The interested reader can complement the information sketched below with Ref. \cite{BLHA}. The interface is based on a two phase model and describes the interaction between a Monte Carlo framework (MC) and a One-loop Program (OLP). During the first phase, the two programs agree on the process and settings for the computation and make all necessary preparation work. This communication (sometimes called negotiation) is done using external text files and does not require the MC and the OLP to be linked or run at the same time. This is particularly convenient in the case where the MC and OLP programs are large and complicated programs generating simpler code to be run in the next phase.

In the second stage the OLP is queried by the MC for the value of the virtual one-loop amplitude for the phase-space points. 
Since speed is relevant in this phase, this communication is done using a direct function call. The two programs will need to be linked at this stage. The run-time programs involved in this phase need not be the same as the ones involved in the first stage and can be more light-weight programs dedicated to the process under consideration, which can make them easier to link them together.  

\subsection{Computing aspects} 
This section lists some computing aspects of the computation of NLO corrections. They are mainly based on the experience of the author with the combination of programs BlackHat+Sherpa, but should apply to NLO corrections in general.

To compute the NLO corrections to a given process both the real and virtual corrections have to be computed. The two contributions are usually computed separately, as they are integrated over different phase spaces. Although the computation of the virtual one-loop amplitude is in general more complicated and typically much more time-consuming the CPU time spent for the virtual part is usually smaller than the time spent computing the real part. This is due to the virtual part beeing integrated over a simpler phase-space with one less external particle. The larger phase-space and larger number of subprocesses for the real part seems to out-weight the added complexity of the one-loop matrix elements. 

Since the phase-space integration is done using Monte Carlo methods, the parallelisation of the work is straightforward and can be described using computing jargon as "embarrassingly parallel". Therefore there is in general no need for a sophisticated parallel architecture. 

A common challenge in the computation of one-loop amplitudes is to ensure the numerical accuracy of the answer. It turns out that the methods chosen by the groups working on the different approaches vary. Groups working with the traditional Feynman diagram method tend to use dedicated evaluation paths for the cases in which the numerical accuracy is endangered. On the other hand, the groups working with either the OPP or generalized unitarity approaches use extended precision to avoid the problem. Evaluating the amplitude with higher precision is costly in terms of CPU time, but this cost can be amortized for a large number of evaluations if the higher precision is not necessary too often. The average cost can also be reduced by only recomputing the parts of the amplitude that caused the accuracy problem. The main advantage of this strategy is that no special routines have to be implemented for a very small number of phase-space points. Additionally, if the program is set up to use higher precision, then the accuracy of the algorithm can be easily tested by comparing with a computation with a much higher precision. 

\section{Conclusions}
A lot of progress has been made in the calculation of one-loop virtual amplitudes. This makes the prospect of a fully automated one-loop calculator more realistic. During the last few years many new processes with four particles in the final state have been computed to NLO accuracy \cite{BDDP,bbbb,ttbbCzakon,ttjj,W3j} and more recently progress has been made towards the calculation of processes with five final-state particles \cite{W4j}. Automated programs for the computation of virtual one-loop amplitudes will also be very practical for use with programs such as {\sc MC@NLO} \cite{MCNLO} and {\sc POWHEG} \cite{POWHEG} that aim at matching NLO matrix elements with parton showers.     


\end{document}